\documentclass[12pt,tableofcontents]{iopart}
\usepackage{graphicx,color,ulem}
\usepackage{amssymb,bm,iopams}
\newcommand{\cK}{\mathcal {K}}
\newcommand{\cP}{\mathcal {P}}
\newcommand{\cQ}{\mathcal {Q}}
\newcommand{\cH}{\mathcal {H}}
\newcommand{\cL}{\mathcal {L}}
\newcommand{\cN}{\mathcal {N}}

\newcommand{\bn}{\mathbf {n}}
\newcommand{\av}[1]{\langle #1 \rangle}
\newcommand{\f}{fluctuator }
\newcommand{\fs}{fluctuators }
\renewcommand{\Re}{\mathop{\mathrm{Re}}}

\newcommand{\prl} {Phys. Rev. Lett.~}
\newcommand{\prb} {Phys. Rev. B~}
\newcommand{\pra} {Phys. Rev. A~}
\newcommand{\me}{m_+^{(\mathrm{e})}}
\begin{document}
\review[Decoherence in qubits due to low-frequency noise]{Decoherence
  in qubits due to low-frequency noise}

\author{J Bergli$^1$, Y M Galperin$^{1,2}$, and B L Altshuler$^3$}

\address{$^1$ Department of Physics, University of Oslo, PO Box 1048
  Blindern, 0316 Oslo Norway}
\address{$^2$ A. F. Ioffe Physico-Technical Institute of Russian
  Academy of Sciences, 194021 St.~Petersburg, Russia}
\address{$^3$ Department of Physics, Columbia University, 538 West
  120th Street, New York, NY10027, USA, and NEC-Laboratories America, Inc., 4 Independence Way, Princeton, NJ 08540, USA}
\ead{jbergli@fys.uio.no}

\begin{abstract}
The efficiency of the future devices for quantum information processing is limited mostly by the finite decoherence rates of the qubits. Recently a substantial progress was achieved in enhancing the time, which a solid-state qubit demonstrates a coherent dynamics. This progress is based mostly on a successful isolation of the qubits from external decoherence sources. Under these conditions the material-inherent sources of noise start to play a crucial role. In most cases the noise that quantum device demonstrate has $1/f $ spectrum. This suggests that the environment that destroys the phase coherence of the qubit can be thought of as a system of two-state fluctuators, which experience random hops between their states. In this short review we discuss the current state of the theory of the decoherence due to the qubit interaction with the fluctuators. We describe the effect of such an environment on different protocols of the qubit manipulations - free induction and echo signal. It turns out that in many important cases the noise produced by the fluctuators is non-Gaussian. Consequently the results of the interaction of the qubit with the fluctuators are not determined by the pair correlation function only.

We describe the effect of the fluctuators using so-called spin-fluctuator model.  Being quite realistic this model allows one to evaluate the qubit dynamics in the presence of one fluctuator exactly. This solution is found, and its features, including non-Gaussian effects are analyzed in details. We extend this consideration for the systems of large number of fluctuators, which interact with the qubit and lead to the $1/f$ noise. We discuss existing experiments on the Josephson qubit manipulation and try to identify non-Gaussian behavior.
\end{abstract}
\submitto{\NJP}

\maketitle

\section{Introduction}

Coherence in quantum solid state devices inevitably suffers from fluctuations due to their environment. In particular, rearrangement of electrons between traps in the insulating regions of the device, as well as stray flux tubes causes pronounced fluctuations in many quantum devices. At low frequencies part of these fluctuations typically has a $1/f$ spectrum and is referred to as $1/f$ noise. Such noise is generic for all disordered materials (for a review see, e.\,g.,~\cite{Dutta81,Weissman88}), it is also common in single-electron and other tunneling devices, see, e.\,g.,~\cite{Zorin96}. Recent experiments~\cite{Astafiev04,Astafiev06} on Josephson qubits indicated that charged impurities may also be responsible for noise.
Low frequency noise is specifically harmful since it is difficult to filter it out by finite band filters.

One of the most common sources of low-frequency noise is the
rearrangement of dynamic two-state defects, {\em fluctuators},
see, e.~g., the book~\cite{Kogan1996} and references therein.
Random switching of a fluctuator  between its two metastable
states (1 and 2)  produces random telegraph noise. The process
is characterized by the switching rates $\gamma_{12}$ and
$\gamma_{21}$ for the transitions $1\to 2$ and $2\to 1$. Only the
fluctuators with energy splitting $E$ less than
temperature, $T$, contribute to the dephasing since the
fluctuators with large level splitting are frozen in their ground
states (we measure temperature in the energy units).  As long as $E<T$ the rates $\gamma_{12}$ and $\gamma_{21}$
are close in magnitude, and without loss of
generality one can assume that $\gamma_{12} \approx
\gamma_{21} \equiv \gamma$.\, i.\,e., the fluctuations can be
described as a \textit{random telegraph process} (RTP),for reviews
see~\cite{Kogan1996,RTP}. A set of random telegraph fluctuators with exponentially broad distribution of relaxation rates, $\gamma$, produces noise with $1/f$ power spectrum at $\gamma_{\min} \ll \omega=2\pi f \ll \gamma_0$. Here $\gamma_{\min}$ is the switching rate of the ``slowest'' fluctuator while $\gamma_0$ is the maximal switching rate for the energy difference between the fluctuators's metastable states equal to temperature.
Random telegraph noise has been observed in numerous nanodevices based both on semiconductors, normal metals, and
superconductors~\cite{RTN}.

Specific features observed in recent experiments~\cite{Simmonds04}
in Josephson phase qubits were interpreted in terms of resonant interaction of the qubit with two-level impurities. These
experiments, as well as results~\cite{Ithier05} in a superconducting
quantum circuit (quantronium), stimulated a very important
attempt~\cite{Shnirman05} to establish a relation between the contributions of two-level fluctuators with low and high frequencies. That has become possible because both of the frequency domains contribute to the dephasing time and the energy relaxation time, respectively. However, this relation was significantly based on the assumption that the statistics of the fluctuations of the qubit parameters  are Gaussian. In our view, this assumption is not obvious, and the main aim of this short review is to discuss he applicability range of the Gaussian approximation, as well as the deviations from the Gaussian approximation in connection with the problem of qubit dephasing.  For this purpose, we will use a simple classical model within which one can evaluate exactly the qubit response to typical manipulation protocols. This model if often referred to as the \textit{spin-fluctuator} (SF) model in similarity with the widely used and sometimes overused
spin-boson model. According to the SF model, the quantum system -- qubit -- interacts with a set of two-level entities. The latter stochastically fluctuate between their states due to interaction with a thermal bath, which may be not directly coupled with  the qubit.
 Since we are interested in the low-frequency noise generated by
these switches the latter can be considered as classical. Consequently, the system qubit$+$fluctuators can be described by relatively simple stochastic differential equations, which in many cases can be exactly solved. In particular, many results can be just borrowed from much earlier papers on
magnetic resonance~\cite{KlauderAnderson,HuWalker}, on spectral diffusion in glasses~\cite{BlackHalperin}, as well as works on single-molecule spectroscopy~\cite{SMS}.

The SF model has previously been used for the description of effects of noise in various systems~\cite{Ludviksson1984,Kogan1984,Kozub1984,Galperin1991,Galperin1994,Hessling1995}
and was recently applied to analysis of decoherence in charge
qubits~\cite{Paladino2001,Paladino2003,Galperin2004,Falci2003,Falci2004,Falci2005,Galperin2006,Martin2006,Bergli2006}.
Various quantum and non-Markovian aspects of the model were addressed in Ref.~\cite{DiVincenzo2005}.  These studies demonstrated, in
particular, that the SF model is suitable for the study of
non-Gaussian effects and that the latter may be essential in certain
situations.

In this short review we will address non-Gaussian effects in
decoherence of Josephson qubits using the echo signal as an
example. Resonant interaction between the qubit and fluctuators
discussed in~\cite{Simmonds04,Galperin05a,Paladino08} will be left out. The paper is organized as follows. In Sec.~\ref{Gaussian} we briefly describe decoherence in the Gaussian approximation. The spin-fluctuator model is described in Sec.~\ref{SF-model}. Section~\ref{experiments} is aimed at discussion of relevance of the SF model to existing experiments, while current microscopic understanding of the two-level fluctuators is briefly reviewed in Sec.~\ref{microscopic}.

\section{Gaussian decoherence} \label{Gaussian}
A qubit is described by the generic Hamiltonian of a pseudospin 1/2 in a ``magnetic field'' $\mathbf{B}$, which can be time-dependent:
\begin{equation}\label{eq:q-H00}
  \cH_q= \frac{1}{2}{\bf B}\cdot \bm{\sigma}
\end{equation}
where $\sigma_i$ are the Pauli matrices.  It is well known that any state vector,  $|\Psi\rangle$, of the qubit determines the Bloch vector ${\bf  M}$ through the density matrix
\begin{equation} \label{eq:rho}
  \rho = |\Psi\rangle\langle\Psi|=
\frac{1}{2} ({\bf 1} + {\bf M}\cdot \bm{\sigma})\, .
\end{equation}
 The Schr\"odinger equation turns out to be equivalent to the precession equation for the Bloch vector:
\begin{equation}\label{eq:precession}
\dot{\bf M}={\bf B}\times{\bf M}.
\end{equation}
The problem of decoherence arises when the ``magnetic field''
 is a sum of a controlled part ${\bf B}_0$ and a fluctuating part ${\bf b}(t)$ which represents the noise, i.\,e.,
is a stochastic process determined by its statistical properties, ${\bf B} = {\bf B}_0 + {\bf b}(t)$. The controlled part ${\bf B}_0$
is not purely static - to manipulate the qubit one has to apply certain high-frequency pulses of $\mathbf{B}_0$ in addition to the static fields applied between manipulation steps. In this paper we will always assume that the manipulation pulses are short enough and neglect the decoherence during the
pulses. Therefore we need to consider only the effect of noise in the presence of static ${\bf B}_0$.

\begin{figure}
\centerline{
\includegraphics[width=5cm]{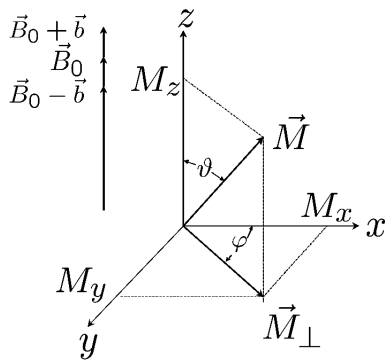}
}
\caption{Bloch vector in the rotating frame of reference. \label{fig_a}}
\end{figure}

We will consider only the case where ${\bf b}\parallel{\bf B_0}$ and let the $z$-axis lie along the common direction of $\mathbf{B}_0$ and $\mathbf{b}$, see Fig.~\ref{fig_a}.
 This situation is called \textit{pure dephasing} because $z$-component of the Bloch vector, $M_z$, is conserved during the process. As long as the time evolution of $\mathbf{M}$ is governed by Eq.~(\ref{eq:precession}), the length $|{\bf M}|$ is also conserved, while the length $|\langle{\bf M}\rangle|$ of  the vector ${\bf M}$ averaged over the stochastic process ${\bf b}$ decays. Description of  this decay is the main objective of the decoherence theory. In the case of pure dephasing, this will be the decay of the components $M_x$ and $M_y$. It is convenient to introduce a complex combination $m_+= (M_x+iM_y)/\sqrt{M_x^2+M_y^2}$. Equation~(\ref{eq:precession}) can be written in terms of $m_+$ as
\[
 \dot{m_+} = iB \, m_+ \quad\mbox{with solution}\quad m_+(t) = e^{\int_0^tB(t')dt'}m_+(0).
\]

This solution has to be averaged over the stochastic process ${\bf b}(t)$. We define the  phase $\phi (t)$ accumulated by $m_+$ as a sum of a regular, $\phi_0$ and stochastic, $\varphi$, parts:
\begin{equation}\label{eq:phi}
 \phi (t) =  \int_0^tdt'B(t') = B_0t+ \int_0^tdt'b(t')= \phi_0 (t)+\varphi (t)\, ,
\end{equation}
and obtain for average value of $m_+$
\[
 \langle m_+(t)\rangle = \langle e^{i\phi}\rangle m_+(0)= e^{i\phi_0}\langle
 e^{i\varphi}\rangle m_+(0).
\]
The Gaussian approximation is based on the assumption that the probability distribution of the phase $\varphi$ is Gaussian:
\begin{equation}\label{eq:gaussdist}
 p(\varphi) = \frac{1}{\sqrt{2\pi\langle\varphi^2\rangle}}\,
   e^{-\frac{\varphi^2}{2\langle\varphi^2\rangle}}\, .
\end{equation}
As usual, this assumption is to be justified on the basis of the central limit theorem. The stochastic phase $\varphi$ is the integral of the random process
\begin{equation} \label{eq:varphi}
v(t) \equiv b(t)\, .
\end{equation}
The Bloch vector precesses around the $z$-axis with the angular velocity that has random modulation $v(t)$ In the Gaussian approximation the only relevant statistical characteristics of $v(t)$ is the correlation function $\langle v(t_1)v(t_2)\rangle =W(|t_1-t_2|)$ (we assume that $v(t)$ is a stationary random process). This function $W(\tau)$ vanishes at $\tau \to \infty$ and the scale of this decay is the correlation time, $\tau_c$. If the time of integration, $t$, is much longer than $\tau_c$ the random phase $\varphi(t)$ is a sum of many uncorrelated contributions. The central limit theorem implies that  such a sum has Gaussian distribution, independently of the details of the process. This is our initial understanding: the Gaussian approximation becomes valid as soon as $t$ exceeds the correlation time of the noise. Below we will further discuss this conclusion. It follows from Eq.~(\ref{eq:gaussdist}) that
\begin{equation}\label{eq:avesp}
    \langle e^{i\varphi}\rangle = \int d\varphi \,  p(\varphi) e^{i\varphi} =
    e^{-\frac{1}{2}\langle\varphi^2\rangle}
\end{equation}
with
\begin{equation}\label{eq:avepx1}
    \langle\varphi^2\rangle = \int_0^tdt_1\int_0^tdt_2 \, \av{v(t_1)v(t_2)}= \int_0^tdt_1\int_0^tdt_2 \, W(|t_1-t_2|)\, .
\end{equation}
The correlator $W(\tau) =\av{v(t)v(t+\tau)}$ is usually represented by its Fourier transform --
the power spectrum of the noise, $S(\omega)$:
\begin{equation}\label{eq:nps}
 S(\omega) = \frac{1}{\pi}\int_0^\infty dt \,W(t)\, \cos\omega t\,  .
\end{equation}
Using Eqs.~(\ref{eq:phi}-\ref{eq:nps}) we obtain
\begin{equation}\label{eq:theta2}
 \langle\varphi^2\rangle = 4\int_{-\infty}^\infty d\omega\,
  \frac{\sin^2\frac{\omega t}{2}}{\omega^2}\, S(\omega) \, .
\end{equation}
For large $t$ the identity
$$\lim_{a\to\infty}\frac{\sin^2ax}{\pi a x^2} = \delta(x)$$
 implies that $\langle\varphi^2\rangle \rightarrow 2\pi t \, S(0)$
 and thus
\begin{equation}\label{eq:gaussdec}
 \langle e^{i\varphi}\rangle = e^{-t/T_2}, \qquad T_2^{-1} = \pi S(0).
\end{equation}
Therefore, the Gaussian approximation leads to the exponential decay of the signal at large times, the decrement being given by the noise power at zero frequency.

\subsection*{Gaussian approximation: Echo experiments}

In the following we will discuss the Bloch vector dynamics in the rotating frame of reference where $\phi_0=0$ and $m_+(t)=e^{i\varphi (t)}$. The time dependence of $\av{m_+}=\av{e^{i\varphi}}$  characterizes decay of the so-called \textit{free induction} signal~\cite{free-ind}. In order to extract it in qubit experiments, one has to average over many repetitions of the same qubit operation. Even in setups that allow single shot measurements \cite{AstafievSS}
each repetition gives one of the two qubit states as the outcome. Only by averaging over many repeated runs can one see the decay of the average as described by the free induction signal. The problem
with this is that the environment has time to change its state between the repetitions, and thus we average not only over the stochastic dynamics of the environment during the time evolution of the qubit, but  over the initial states of the environment as well. As a result, the free induction signal decays even if the environment is too slow to rearrange during the operation time. This is an analog of the inhomogeneous broadening of spectral lines in magnetic resonance experiments. This analogy also suggests ways to eliminate the suppression of the signal by the dispersion of the initial conditions. One can use well known echo technique(see, e.\,g.,~\cite{Mims})when the system is subject to a short manipulation pulse (so-called $\pi$-pulse) with duration $\tau_1$ at the time $\tau_{12}$. The duration $\tau_1$ of the pulse is chosen to be such that it switches the two states of the qubit. This is equivalent to reversing the direction of the  Bloch vector and thus effectively reversing the time evolution after the pulse as compared with the initial one. As a result, the effect of any static
field is canceled and decay of the echo signal is determined only by the \textit{dynamics} of the environment. The decay of 2-pulse echo can be expressed as $\av{\me(2\tau_{12})}$~\cite{Mims}, where
\begin{equation}
  \label{eq:016}
  \av{\me (t)} \equiv \av{e^{i\psi(t)}}\, , \quad
\psi(t)=\left(\int_0^{\tau_{12}}-\int_{\tau_{12}}^t
\right)v(t^\prime) \, d t^\prime \, .
\end{equation}
Finite correlation time of $v(t)$ again leads to the Gaussian distribution of $\psi (t)$ at large enough $t$ with
\begin{equation}\label{eq:theta2echo}
 \langle\psi^2(2\tau_{12})\rangle = 16\int_{-\infty}^\infty d\omega
  \frac{\sin^4\frac{\omega \tau_{12}}{2}}{\omega^2} S(\omega).
\end{equation}
This variance can be much smaller than $\av{\varphi^2}$, Eq.~(\ref{eq:theta2}), if $S(\omega)$ is singular at $\omega \to 0$.

\section{Non-Gaussian decoherence: Spin-fluctuator
  model}\label{SF-model}

We have discussed how to calculate the decoherence of a qubit in the
Gaussian approximation. The only statistical characteristic of the noise is the time correlation function, $W(t)$, or equivalently the  power spectrum, $S(\omega)$. Now let us consider situations when this approximation is not valid. In such situations knowledge of only the noise power spectrum is not sufficient: noise sources with identical power spectra can have different decohering effect on the qubit. Thus, it is necessary to specify the  model for the noise source in more detail. In the following we will  use a random telegraph process as a model noise source. We will first analyze individual random telegraph processes  and show how $1/f$ noise appears as a result of averaging over a suitable  ensemble of these processes. We will describe the decoherence due to one telegraph process, and finally extend this discussion to decoherence by averaging over  ensembles of telegraph processes.

\subsection{Random telegraph processes}

Consider a stochastic function $\chi(t)$, which
at any time takes the values $\chi(t)=\pm1$~\cite{RTP}.  It is thus
suitable for  describing a system that can find itself in one of the
two stable states, 1 and 2,  and once in a while makes a
switch between them. The switchings are assumed to be uncorrelated random  events with rates $\gamma_{12}$ and $\gamma_{21}$, which in principle  can be different. Here we will limit ourselves to symmetric telegraph process: $\gamma_{12}=\gamma_{21}=\gamma$. The extension to the general case is straightforward, see, e.\,g.,~\cite{Itakura03}. The number $k$ of switches that the fluctuator experiences within a time $t$  follows a Poisson distribution
\begin{equation}\label{eq:poisson} P_k=\frac{(\gamma
t)^k}{k!}e^{-\gamma t} \,.
\end{equation}
The number of switches, $k$, determines the number of times
the function $\chi(t)$ changes its sign contributing $(-1)^k$ to the correlation function, $C(t) \equiv \langle \chi(t)\chi(0)\rangle$. Therefore
\begin{equation}
    C(t) %\equiv \langle \chi(t)\chi(0)\rangle
   % = \sum_{k=0}^\infty \chi_k(t)\chi(0)P_k
    = e^{-\gamma t}\sum_{k=0}^\infty (-1)^k\frac{(\gamma t)^k}{k!}
    = e^{-2\gamma t}\, , \quad t \ge 0\, .
\end{equation}
The random telegraph process results in fluctuating field ${\bf v}(t)={\bf v} \chi(t)$ applied to the qubit. The magnitude of this field, $v= |{\bf v}|$, together with the switching rate, $\gamma$, characterizes the fluctuator. Using (\ref{eq:nps}) we find power spectrum of the noise generated by $i$-th fluctuator:
\begin{equation}\label{eq:srtp}
 S_i(\omega) = v^2_i\cdot \frac{1}{\pi}
 \frac{2\gamma_i}{(2\gamma_i)^2+\omega^2}\, .
\end{equation}

\paragraph{Generating $1/f$ noise by sets of telegraph processes. --}
Let us consider a combination of a few statistically
independent telegraph processes. Since the effective fields are all parallel to the $z$-axis we can characterize $i$-th fluctuator by a coupling strength $v_i$ and a switching rate $\gamma_i$ so that the total magnitude of the fluctuating field is given by $\sum_i v_i\chi_i(t)$. The total noise spectrum  $S(\omega)$ equals to a sum
$\sum_i S_i(\omega)$. If the number of fluctuators interacting with the qubit is large the fluctuating field can be written as a convolution of $S_i(\omega)$, Eq.~(\ref{eq:srtp}), with the distribution $P(v,\gamma) $ of the parameters $v$ and $\gamma$,
 \begin{equation} \label{eq:avnps}
  S(\omega)=\frac{1}{\pi}\int v^2 dv\int d\gamma\,  \cP (v, \gamma)\,
  \frac{2 \gamma \, }{\omega^2+(2\gamma)^2}\, .
\end{equation}
To obtain the $1/\omega$ low-frequency behavior of the power spectrum (\ref{eq:avnps}) one has to assume that the distribution function $P(v,\gamma)$ behaves as $1/\gamma$ at small $\gamma$, i.\,e.,
 \begin{equation}\label{eq:avnps1}
 \cP(v,\gamma)\vert_{\gamma \to 0} = \cP(v)/\gamma\, .
 \end{equation}
In this case,
\begin{equation}\label{eq:somega0}
    S(\omega)=\av{v^2}/\omega\, , \quad \av{v^2}=\int_0^\infty v^2\cP(v)\, dv\,.
\end{equation}
It turns out that the distribution (\ref{eq:avnps1}), which corresponds to a uniform distribution of $\log \gamma$, follows naturally from a simple model. In this commonly used model the role of a bistable fluctuator is played by a particle confined by a double well potential~\cite{TLS}; each well has one state to host the particle, and the tunneling between these states is possible. This two level system (TLS) describes a broad class of fluctuators: the particle can be a generalized particle moving in a generalized space. Hamiltonian of a TLS with tunneling can be written as
\begin{equation}
  \label{eq:hf} \cH_F=\frac{1}{2}\left(\Delta \tau_z+\Lambda
\tau_x\right)
\end{equation}
where $\tau_i$ are the Pauli matrices (in contrast with the Pauli matrices $\sigma_i$ that act in the qubit Hilbert space). Each TLS is characterized by two parameters -- diagonal splitting, $\Delta$, and tunneling matrix element, $\Lambda=\hbar \omega_0 e^{-\lambda}$. Here $ \omega_0$ is a typical frequency of the classical motion of the ``particle'' inside each of the two wells. The logarithm of the switching rate $\gamma$ is proportional to the dimensionless tunneling integral $\lambda$. A natural assumption that $\lambda$ is uniformly distributed in the TLS ensemble leads to the distribution (\ref{eq:avnps1}) for the switching rate $\gamma$.

The environment is usually modeled as a thermal bath, which can
represent a phonon field as well as, e.\,g., electron-hole pairs
in the conducting part of the system.
Fluctuations in the environment effect the fluctuator through either $\Delta$ or $\Lambda$, Eq.~(\ref{eq:hf}). Assuming that the modulations of the diagonal splitting $\Delta$ are most important we can describe the interaction of the environment with the qubit as
 \begin{equation}
  \label{eq:hfe0}
  \cH_{F-\mbox{env}}=g' \hat{c}\tau_z,
\end{equation}
where $\hat{c}$ is an operator in the Hilbert space of the environment depending on the concrete interaction mechanism.
It is convenient to diagonalize $\cH_F$ (\ref{eq:hf}) by rotating of the fluctuator Hilbert space. Then
$$\cH_F=\frac{E}{2} \tau_z\, , \quad E=\sqrt{\Delta^2+\Lambda^2}\, . $$
 Keeping the notation $\tau_i$ for the
Pauli matrices representing the fluctuator in the rotated basis we write the interaction Hamiltonian (\ref{eq:hfe0}) as
\begin{equation}
  \label{eq:hfe}
  \cH_{F-\mbox{env}}=g' \hat{c}\left(\frac{\Delta}{E}\tau_z -
  \frac{\Lambda}{E}\tau_x\right)\,  .
\end{equation}
The factor $(\Lambda /E)^2$
appears in the inter-level transition rate \cite{gamma,Black}:
\begin{equation}
  \label{eq:gamma}
  \gamma (E,\Lambda)=(\Lambda/E)^2\, \gamma_0 (E)\, .
\end{equation}
Here the quantity $\gamma_0 (E)$ has the meaning of the {\em maximal} relaxation rate for fluctuators with a given energy splitting, $E$.

The parameters $\Delta$ and $\lambda$ are usually supposed to
be uniformly distributed over intervals much larger than those
important for low-temperature kinetics~\cite{gamma,dist2}, so we can
write their probability distribution as
\begin{equation} \label{eq:pdist1}
\cP(\lambda,\Delta)=P_0
\end{equation}
(leaving aside the problem of normalization, which will be fixed
later). It is convenient to characterize each fluctuator by two parameters -- the energy spacing between the levels, $E=\sqrt{\Delta^2+\Lambda^2}$, and $\theta$ -- determined through the relations $\Delta = E\cos \theta$, $\Lambda =E\sin \theta$ ($\theta \le \pi/2$). The mutual distribution function of these parameters can we written as~\cite{Laikhtman}
\begin{equation}
  \label{eq:012ab}
  \cP (E,\theta)=\frac{P_0}{\sin \theta}\, , \quad 0 \le \theta
  \le \frac{\pi}{2}\, .
\end{equation}
Relaxation rates $\gamma$ for the fluctuators with a given spacing $E$ are distributed according to
\begin{equation}
  \label{eq:012a}
   \cP (E,\gamma)=\frac{P_0}{2\gamma} \left[1-\frac{\gamma}{\gamma_0(E)}\right]^{-1/2}\, ,
     \quad \gamma_{\min}(E) \le \gamma  \le \gamma_0 (E)\, .
\end{equation}
To normalize this distribution one has to cut it off  at small relaxation rates at a minimal value $\gamma_{\min}(E)$. The distribution (\ref{eq:012a}) has Eq.~(\ref{eq:avnps1}) as its limit at $\gamma \ll \gamma_0$ and thus leads to $1/f$ noise in the frequency domain $\gamma_{\min} \ll \omega \ll \gamma_0$. The coupling, $v$, between the qubit and the fluctuator  depends on the parameters $E$ and $\theta$ of the fluctuator, see Sec.~\ref{ensemble-average} for more details. Since only the fluctuators with $E \lesssim T$ are important and both $v$ and $\gamma_0$ are smooth functions of $E$ one can  use the values of $v$ and $\gamma_0$ at $E = T$. Below we restrict ourselves by this approximation which allows us to perform integration over the energy spacings of the fluctuator states. In particular, integrating Eq.~(\ref{eq:012a}) over $E$ and $\gamma$ we connect $P_0$ with the total number, $N_T$, of thermally excited fluctuators:
 \begin{equation}
  \label{eq:013a}
  P_0 =\frac{\cN_T}{T \cL}\, , \quad \cL \equiv \ln \frac{\gamma_0 (T)}{\gamma_{\min}(T)}\, .
\end{equation}
The integral expression (\ref{eq:avnps}) for the noise power spectrum is valid provided that  $P_0T \gg 1$, i.\,e., $\cN_T \gg \cL$.
Substituting Eq.~(\ref{eq:013a}) into Eq.~(\ref{eq:avnps}) we obtain the estimate:
\begin{equation}
  \label{eq:013}
  S(\omega)\sim \av{v^2} P_0T \left\{
\begin{array}{ccc}
1/\omega&,&\gamma_{\min} \ll \omega \ll \gamma_0 \, ; \\
\gamma_0/\omega^2 &,& \omega \gg \gamma_0\, . \end{array} \right.
\end{equation}
Therefore, the noise in the SF model indeed has the $1/f$ noise power spectrum at low enough frequencies. The crossover from $\omega^{-1}$ to $\omega^{-2}$ behavior at $\omega \sim \gamma_0$ manifests the cutoff in the distribution (\ref{eq:012a}) at high switching rates.

\subsection{Decoherence by a single random telegraph process}
\paragraph{Master equations. --}
Having described the properties of random telegraph noise we are
now ready to discuss how it effects a qubit.
Let us turn to Eq.~(\ref{eq:precession}) with a single telegraph process as the noise source.

We assume that the fluctuator does not feel any feedback from the qubit and thus the random telegraph function $\chi (t)$ equals to +1 or -1 with the probability 1/2 regardless to the direction of the Bloch vector $\mathbf{M}$. Consider the dynamics of the Bloch vector in the rotating frame of reference during a time interval $t,t+\tau$, which is small as compared to the inverse switching rate $1/\gamma$, so that the  fluctuator changes its state with a small probability, $\gamma \tau \ll 1$ while the probability to switch more than once is negligible.

Let us split the probability to find the angle $\varphi$ at time $t$, $p(\varphi,t)$, into partial probabilities to arrive at the angle $\varphi$ due to rotation around the vector $\mathbf{b}$ and $-\mathbf{b}$, respectively:
\begin {equation}\label{eq:p01}
p(\varphi,t)=p_+(\varphi,t)+p_+(\varphi,t)\, .
\end{equation}
Denoting as $\cP(\tau)$ the probability  for a fluctuator \textit{not} to switch during  the time interval $(t,t+\tau)$ and the probability to switch once as $\cQ(\tau)$ one can express $p_+(\varphi,t+\tau)$ through $p_+(\varphi,t)$ and $p_-(\varphi,t)$:
\begin{equation}\label{eq:me01} \fl
p_+(\varphi,t+\tau)= \cP(\tau) p_+(\varphi +v\tau,t)+ \int_t^{t+\tau}dt_1\left\{ \dot{\cQ}(t_1-t)p_-[\varphi-v(t+\tau- t_1),t]\right\}\, .
\end{equation}
A similar equation can be written for $p_-(\varphi,t+\tau)$:
\begin{equation}\label{eq:me02} \fl
p_-(\varphi,t+\tau)= \cP(\tau) p_-(\varphi -v\tau,t)+ \int_t^{t+\tau}dt_1\left\{ \dot{\cQ}(t_1-t)p_+[\varphi+v(t+\tau- t_1),t]\right\}\, .
\end{equation}
When $\tau \ll \gamma^{-1}$ we can indeed neglect multiple switchings, and expand Eq.~(\ref{eq:poisson}) as
$\cP(t)=P_0=1-\gamma \tau$, $\cQ (t) =P_1=\gamma \tau\, .$
Differentiating Eqs.~(\ref{eq:me01}) and (\ref{eq:me02}) over $\tau$ we arrive at the set of equations:
\begin{eqnarray} \label{eq:me02a}
  \dot{p}_+ &=& -\gamma p_++\gamma p_-+v\partial_\varphi p_+\, , \nonumber\\
  \dot{p}_- &=& -\gamma p_-+\gamma p_+-v\partial_\varphi p_-\, ,
\end{eqnarray}
$\partial_\varphi p_\pm \equiv \partial p_\pm /\partial \varphi$. Equations (\ref{eq:me02a}) can be combined to one second-order differential equation for $p(\varphi,t)$, Eq.~(\ref{eq:p01}),
\begin{equation} \label{eq:tmp003}
\ddot{p}+2\gamma \dot{p}=v^2\partial^2_\varphi p\, ,
\end{equation}
which is known as the telegraph equation. We can always choose $x$-direction in such a way that $\varphi=0$ at $t=0$. Thus $p(\varphi,0)=\delta(\varphi)$. The second initial condition to Eq.~(\ref{eq:tmp003}), $\dot{p}(\varphi,0)=\pm 2 v\partial_\varphi p(\varphi,0)$, follows directly from the integral equation (\ref{eq:me01}). The sign depends on the initial state of the fluctuator. After averaging over the fluctuator's initial state,
$\dot{p}(\varphi,0)=0$.

To evaluate the decoherence for this model we multiply Eq.~(\ref{eq:tmp003}) by $m_+ =e^{i\varphi}$ and integrate over $\varphi$ to show that the mean value of $m_+$,
\[
 \langle m_+\rangle =  \langle e^{i\varphi}\rangle = \int d\varphi \, p(\varphi)\,
e^{i\varphi},
\]
satisfies the equation
\begin{equation}\label{eq:tmp005a}
 \langle \ddot{m}_+\rangle +
  2\gamma \langle \dot{m}_+\rangle  = -v^2\langle m_+\rangle\,.
\end{equation}
The initial condition for this equation is $m_+(0)=1$ since $\varphi (0)=0$. The second initial condition, $\dot{m}_+(0)=0$, follows from the boundary condition $\dot{p}(\phi,0)=0$. The solution of Eq.~(\ref{eq:tmp005a}) with these initial conditions,
\begin{equation}\label{eq:rtpdec}
 \langle m_+\rangle  = \frac{e^{-\gamma t}}{2\mu}\left[(\mu+1)e^{\gamma\mu t}
  + (\mu-1)e^{-\gamma\mu t}\right], \quad \mu \equiv
\sqrt{1-\frac{v^2}{\gamma^2}}\, ,
\end{equation}
describes decoherence of a qubit due to a single random telegraph process given by the coupling strength $v$ and the switching rate $\gamma$.

According to (\ref{eq:rtpdec}), the free induction signal demonstrates qualitatively different behavior for large and small ratios $v/\gamma$. At $v \gg \gamma$ one can consider the qubit as a quantum
  system experiencing beatings between the states with
  different splittings,  $\mathbf{B}_0\pm \mathbf{b}$, the width of
  these states being $\gamma$. In the
  opposite limiting case, $v \ll \gamma$, the inter-level splitting  is self-averaged to a certain  value, the width being $v^2/2\gamma$. This situation was extensively discussed in connection with the magnetic resonance and is known as the \textit{motional narrowing of spectral   lines}~\cite{KlauderAnderson}. The two types of behavior will be
  discussed in more detail in Sec.~\ref{experiments}.

\begin{figure}[h]
\begin{center}
\includegraphics[width=8.0cm]{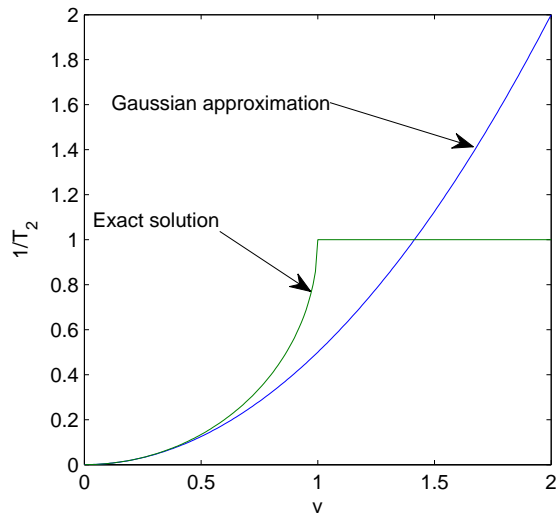}
\end{center}
\caption{
Comparison of the decoherence rate $T_2^{-1}$ for a single
random telegraph process and the corresponding Gaussian approximation.
\label{fig:t2}}
\end{figure}

\paragraph{Comparison to the Gaussian approximation. --}
 Let us now compare Eq.~(\ref{eq:rtpdec}) with the result (\ref{eq:gaussdec}) of the
Gaussian approximation. Substituting (\ref{eq:srtp}) for the noise power spectrum one obtains
\begin{equation}\label{eq:t2g}
 \frac{1}{T_2^{(G)}} = \frac{v^2}{2\gamma} \, .
\end{equation}
When discussing the Gaussian approximation we argued that it should be valid for times longer than the correlation time of the noise, which for the random telegraph process is $(2\gamma)^{-1}$. Expanding our  Eq.~(\ref{eq:rtpdec}) at long times we find that it indeed decays exponentially (or the oscillations decay exponentially in the case $v>\gamma$). However the rate of the decay is parametrically different from Eq.~(\ref{eq:t2g}):
\begin{equation}\label{eq:t2rtp}
\frac{1}{T_2} = \gamma-\gamma\Re\sqrt{1-\frac{v^2}{\gamma^2}} \, .
\end{equation}
It is easy to check that in the limit $v\ll\gamma$ Eq.~(\ref{eq:t2rtp})coincides with the Gaussian result.
Shown in Fig.~\ref{fig:t2} are equations (\ref{eq:t2g})
and (\ref{eq:t2rtp}) as functions of $v$ for fixed $\gamma=1$.
We see that the Gaussian approximation is only valid in the limit
$v\ll\gamma$. Apparently this conclusion is in a contradiction with the previous discussion based on the central limit theorem. It looked convincing that according to this theorem $p(\varphi,t)$ \textit{always} tends to a Gaussian distribution with time-dependent variance provided that the time exceeds the correlation time of the noise.

\begin{figure}[h]
\begin{center}
\includegraphics[width=8.0cm]{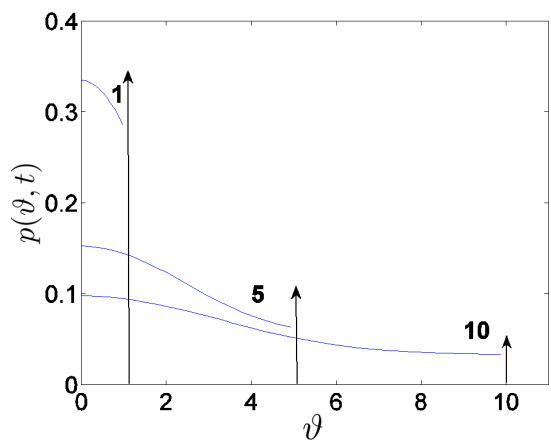}
\end{center}
\caption{
The distribution function (\ref{eq:p}) for $t=1$, 5 and 10 (only the part
for positive $\varphi$ is shown, the function is symmetric). $v=\gamma=1$.
The arrows represent the delta-functions (not to scale).
\label{fig:p}}
\end{figure}

To resolve this apparent contradiction let us analyze the shape of the distribution function, $p(\varphi,t)$, which follows from Eq.~(\ref{eq:tmp003}).
The solution of this equation with the boundary conditions
$p(\varphi,0)=\delta(\varphi)$ and $\dot{p}(\varphi,0)=0$  is  \cite{Bergli2006}:\footnote{Note that in Ref.~\cite{Bergli2006} the term with $I_0$ was missing.}
\begin{eqnarray}\label{eq:p}
 p(\varphi,t) &=& \frac{1}{2}e^{-\gamma t}
  \left[\delta(\varphi+vt)+\delta(\varphi-vt)\right]
 +\frac{\gamma}{2v}e^{-\gamma t}\left[\Theta(\varphi+vt)-\Theta(\varphi-vt)\right] \nonumber\\
  &&\times \left[\frac{I_1\left(\gamma t\sqrt{1-(\varphi/vt)^2}\right)}
         {\sqrt{1-(\varphi/vt)^2}}
    + I_0\left(\gamma t\sqrt{1-(\varphi/vt)^2}\right)\right]
  \end{eqnarray}
where $I_v(x)$ is the modified Bessel function and $\theta(x)$ is the Heaviside step function:
 $$\Theta (x)=\left\{\begin{array}{ll}
 1,&x>0\, ,\\ 0,& x<0\, . \end{array}\right. $$
 The distribution function (\ref{eq:p}) (shown in Fig.~\ref{fig:p} for various $t$)
consists of two delta-functions and a central peak.

The delta-functions represent the finite probability for the fluctuator to be in the same state during time $t$. As time increases, the weight of the delta-functions decreases and the central peak broadens. At large times this peak acquires a Gaussian shape. Indeed, at $\gamma t \gg 1$ one can use the asymptotic behavior of the Bessel function
$
 I_v(z) \to (2\pi)^{-1/2}e^z
$,
as $z \to \infty$. For $\varphi\ll vt$ we can also expand $\sqrt{1-(\varphi/vt)^2}$ and convince ourselves that
the central peak in accordance with the central limit theorem is indeed described by the Gaussian distribution (\ref{eq:gaussdist})
with the variance $\av{\varphi^2} = v^2t/\gamma$.
 If the qubit-fluctuator coupling is weak, $v\ll\gamma$,
this Gaussian part of $p(\varphi,t)$ dominates the average $\av{e^{i\varphi}}$ and the Gaussian approximation is valid. Contrarily, for the strong coupling,  $v>\gamma$, the average is
dominated by the delta functions at the ends of the distribution, and the decoherence demonstrates a  pronounced non-Gaussian behavior, even at $t > \gamma^{-1}$.

Unfortunately it is impossible to measure the distribution of $\varphi$ experimentally. Indeed, each experimental shot corresponds to a particular realization of the random external noise and does not yield a particular value of $\varphi$. The reason is in the difference between the qubit that can can be viewed as a pseudospin 1/2 and a classical Bloch vector $\mathbf{M}$. According to Eq.~(\ref{eq:rho}) the components $M_x,M_y,M_z$ of $\mathbf{M}$ are connected with the mean component of the final state of the pseudospin. Therefore, to measure the value of the phase $\varphi$ (argument of $m_+$) that corresponds to a given realization of the noise one would have to repeat the experimental shot with \textit{the same realization of the noise} many times. This is impossible because each time the realization of noise is different. Therefore, the only observable in the decoherence experiments is the average $\av{e^{i\varphi}}$. There is no way to extract more information about the distribution $p(\varphi,t)$ from any experiment with a single qubit.

\subsection
{Averaging over ensembles of random telegraph processes}
\label{ensemble-average}
\paragraph{General expressions. --}
As we have seen, a set of fluctuators characterized by the
distribution function~(\ref{eq:012a}) of relaxation rates provides a
realistic model for $1/f$ noise. Thus, in order to study the decoherence by such noise it is natural to sum contributions of many fluctuators. To perform this procedure we assume that dynamics
of different fluctuators are not correlated, i.\,e., $\langle
\chi_i(t) \chi_j(t')\rangle = \delta_{ij}e^{-2\gamma_i |t-t'|}$.
Under this assumption the average value of the complex momentum,
$\langle m_+\rangle$ is just a product of the partial averages,
$$\av{m_+(t)}=\prod_i\av{m_{+i} (t)}=e^{\sum_i \ln \av{m_{+i}(t)}}\, .$$
Since the logarithm of a product is a self-averaging quantity, it is
natural to approximate the sum of logarithms, $\sum_i \ln \av{m_+(t)}_i$, by its average value,
 $-\cK_m(t)$, where
\begin{equation}
  \label{eq:avlog}
\cK_m(t) \equiv -\overline{
\sum_i \ln\av{m_{+i}(t)}}\, .
\end{equation}
Here bar denotes the average over both
the coupling constants, $v$, of the fluctuators and  their transition rates, $\gamma$. If the
number  $\cN_T$ of thermally excited \fs is large we can replace the
sum $\sum_i \ln \av{m_{+i}}$ by $\cN_T \overline{\ln \av{m_+}}$.
Furthermore, one can employ the Holtsmark procedure~\cite{Chandrasekhar}, i.~e., to replace $\overline{\ln \av{m_+}}$ by $\overline{\av{m_+} -1}$, assuming that each of $\av{m_{+i}}$ is close to $1$. Thus, $\cK_m (t)$ is approximately equal
\begin{equation}  \label{Holtsmark}
\cK_m(t)\approx \cN_T \overline{(
  1-\av{m_+ })}=\int dv\, d\gamma\,
\cP(v,\gamma)\left[1-\av{m_+(v,\gamma|t)}\right]\, .
\end{equation}
Here $\av{m_+(v,\gamma|t)}$ depends on the parameters $v$ and $\gamma$ according to Eq.~(\ref{eq:rtpdec}). The average free
induction signal is then $\overline{\av{m_+(t)}}=\e^{-\cK_m(t)}$.

Analysis of the echo signal is rather similar: one has to replace $\av{m_+(t)}$
taken from Eq.~(\ref{eq:rtpdec}) by~\cite{Laikhtman}
\begin{eqnarray}
  \langle \me (2\tau_{12})\rangle = \frac{e^{-2\gamma\tau_{12}}}{2\mu^2}
  \left[(\mu+1)e^{2\gamma\mu\tau_{12}}-(\mu-1)e^{-2\gamma\mu\tau_{12}}
    - \frac{2v^2}{\gamma^2}\right]
\, .  \label{eq:04}
 \end{eqnarray}

To evaluate the time dependence of either the free induction or the echo signal one has to specify the distribution of the coupling constants $v$. Let us consider each fluctuator as a two-level tunneling system characterized by the Hamiltonian
\begin{equation}
  \label{eq:hfi}
  \cH_F^{(i)}=\frac{1}{2}\left(\Delta_i \tau_z^{(i)}+\Lambda_i
    \tau_x^{(i)}\right)
\end{equation}
where $\bm{\tau}^{(i)}$ is the set of Pauli matrices describing $i$-th two-state fluctuator. The energy splitting for each fluctuator is $E_i=\sqrt{\Delta^2_i+ \Lambda^2_i}$.
 The variation of the energy spacing between the states of the qubit can be cast into the effective Hamiltonian, which (after a rotation similar to that in Eq.~(\ref{eq:hfe})) acquires the form
\begin{equation}
  \label{eq:004b}
 \cH_{qF}=\sum_i v_i \,
 \sigma_z\tau_z^{(i)}\, , \quad  v_i=g(r_i)A(\bn_i)\cos \theta_i
\, .
\end{equation}
Here $\theta_i \equiv \arctan (\Lambda_i/\Delta_i)$,
$\bn_i$ is the direction of elastic or electric dipole moment of $i$-th fluctuator, and $r_i$ is the distance between the qubit  and $i$-th fluctuator. Note that in Eq.~(\ref{eq:004b}) we neglected the
term proportional to $\sigma_z\tau_x$. This can be justified as long as the fluctuator is considered to be a classical system. The functions $A(\bn_i)$ and $g(r_i)$ are not universal.

The coupling constants, $v_i$, determined by Eq.~(\ref{eq:004b}),
contain $\cos \theta_i$ and thus are statistically correlated with
$\theta_i$. It is convenient to introduce an uncorrelated random
coupling parameter, $u_i$ as
\begin{equation}
  \label{eq:coupling}
 u_i=g(r_i)A(\bn_i)\, , \ v_i=u_i\cos \theta_i\, .
\end{equation}
It is safe to assume that direction, $\bn_i$, of a fluctuator is
correlated neither with its distance from the qubit, $r_i$, nor
with the tunneling parameter represented by the variable $\theta_i$. This assumption allows us to replace $A(\bn)$ by its angle average, $ \bar{A} \equiv \langle |A(\bn)|\rangle_\bn$.

\paragraph{Simple model. --}
\begin{figure}[h]
\begin{center}
\includegraphics[width=7.0cm]{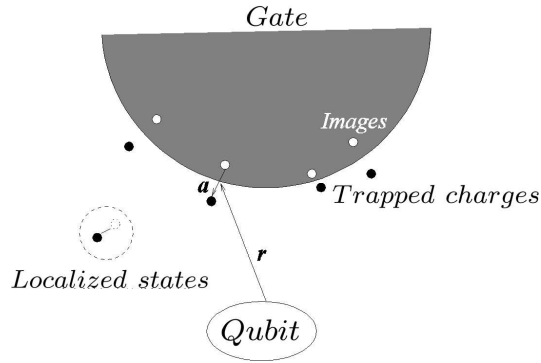}
\end{center}
\caption{Sketch of localized charges near an electrode. Induced image charges
create local dipoles which interact with the qubit.
\label{fig:fig1}}
\end{figure}
The coupling parameter, $g$, decays with distance, $r$, between the qubit and the fluctuator. Usually this decay is algebraic: $g(r)\propto \bar{g}/r^b$. We will distinguish between two cases:
(i) the fluctuators are distributed in three dimensional space (d=3) and
(ii) the fluctuators are located in the vicinity of a two dimensional manifold, e.\,g., in the vicinity of the interface between an insulator and a metal (d=2). Using the distribution (\ref{eq:012a}) of the relaxation rates one can express $\cP(u,\theta)$ as
 \begin{equation}
  \label{eq:06}
\cP (u,\theta)=(\eta \cos \theta)^{d/b}\frac{1 }{\sin\theta}\,\frac{1} {u^{d/b+1}}\, , \quad
\eta  \equiv \frac{\bar{g}}{r_T^b}\, , \ r_T\equiv \frac{a_d}{( P_0
    T)^{1/d}} \,  .
\end{equation}
Here $a_d$ is a $d$-dependent dimensionless constant while $r_T$ is a typical distance between the fluctuators with $E_i\lesssim T$. In the following we will for simplicity assume that
\begin{equation}
\label{eq:mf} r_{\min} \ll r_T \ll r_{\max}\, ,
\end{equation}
 where $r_{\min}$ ($r_{\max}$) are distances between the qubit and the closest (most remote) fluctuator. Under this condition $\eta \propto
T^{b/d}$ is the typical constant of the qubit-fluctuator
coupling. As soon as the inequality (\ref{eq:mf}) is violated the
decoherence starts to depend explicitly on either $r_{\min}$ or
$r_{\max}$, i.~e., becomes sensitive to particular mesoscopic details of the device.

Let us first consider the case when $d=b$, as it is for charged
traps located near the gate electrode (were $b=d=2$ due to the dipole nature of the field produced by a charge and its induced image, see
Fig. \ref{fig:fig1}). In this case one can cast Eq.~(\ref{Holtsmark}) into the form
\begin{equation}
  \label{eq:Hotlsmark1}
  \cK_f(t)=\eta \int \frac{du}{u^2} \int_0^{\pi/2}\! d\theta \tan \theta
  \left[1-f(u\cos \theta, \gamma_0 \sin^2 \theta|t)\right]\, .
\end{equation}
Here $f(v,\gamma|t)$ is equal either to $\av{m_+(v,\gamma|t)}$ or to
$\av{\me (v,\gamma|t)}$, depending on the manipulation
protocol. Equation (\ref{eq:Hotlsmark1}) together with
Eqs.~(\ref{eq:rtpdec}) and (\ref{eq:04}) allows one to analyze
different limiting cases.

In the case of two-pulse echo it is instructive to look at the
asymptotic behaviors of $ \cK_f(t)$. {}From Eq.~(\ref{eq:04}) it follows that
\begin{equation}
   \label{eq:018a} \fl
   1-\av{\me (u\cos \theta, \gamma_0 \sin^2 \theta|t)} \propto
  \left\{\begin{array}{lll}
    t^3 \, \gamma_0\sin \theta  (u \cos \theta)^2   &,&t \ll
    (\gamma_0\sin \theta)^{-1}, \\
    t^2\, (u \cos\theta)^2  &,&   (\gamma_0\sin \theta)^{-1}\ll t \ll u^{-1},\\
    t \,  u \cos \theta   &,&   u^{-1}\ll t.
  \end{array} \right.
\end{equation}
Splitting the regions of integration over $u$ and $\theta$
according the domains (\ref{eq:018a}) of different asymptotic behavior
one obtains
\begin{equation}
  \label{eq:0018b}
  \cK_m(2\tau_{12})\sim \eta  \tau_{12} \min\{\gamma_0 \tau_{12},
  1\}\, .
\end{equation}

The dephasing time (defined for non-exponential decay as the time when $\cK\sim1$)
for the two-pulse echo decay is thus equal to
\begin{equation}
  \label{eq:10a}
  \tau_\varphi =\max\left \{\eta^{-1},(\eta \gamma_0)^{-1/2}\right \}\,.
\end{equation}

The result for $\gamma_0 \tau_{12} \ll 1$ has a clear physical
meaning~\cite{Laikhtman}: the decoherence occurs only provided that at least one of the fluctuators flips. Each flip provides a contribution $\sim \eta t$ to the phase, while $\gamma_0 \tau_{12} \ll 1$ is a probability for a flip during the observation time.  The result for $\gamma_0 \tau_{12}\gg 1$ is less intuitive since in this region the dephasing is non-Markovian, see~\cite{Laikhtman} for more detail.

It is important that at large observation times, $\tau_{12} \gg
\gamma_0^{-1}$, the decoherence is dominated by few  \emph{optimal} \fs. The distance $r_{\scriptsize \textrm{opt}}(T)$, between
the optimal \fs and the qubit is determined by the condition
  \begin{equation}
    \label{eq:aux1}
    v(r_{\scriptsize\textrm{opt}}) \approx \gamma_0(T)\, .
  \end{equation}
Derivation of Eq.~(\ref{eq:aux1} requires a rather tedious analysis
of the expansion~(\ref{eq:018a}) and the integration over $u$ and
$\theta$, see~\cite{Laikhtman,Galperin2004} for more detail. This estimate emerges naturally from the behavior of the decoherence in the limiting cases $v \gg \gamma$ and $v \ll \gamma$. For strong coupling the fluctuators are slow and the qubit's behavior is determined by quantum beatings between the states with $E\pm v$. Accordingly, the decoherence rate is of he order of $\gamma$. In the opposite case, as we already discussed, the decoherence rate is $\propto v^2/\gamma$. Matching these two limiting cases one arrives at the estimate~(\ref{eq:aux1}).

What if $d \ne b$? If the coupling decays as $1/r^b$ and the \fs are distributed in a  $d$-dimen\-sio\-nal space, then
$r^{d-1}\, dr \to \cP(v) \propto v^{-(1+d/b)}$. As a result, at $d \le b$ the decoherence is controlled by optimal fluctuators located at the distance $r_{\scriptsize \textrm{opt}}$
\emph{provided that they exist}. At $d >  b$ the decoherence at large time is determined by most remote fluctuators with $v=v_{\min}$.
If $d \le b$, but the closest \f has $v_{\max} \ll \gamma_0$, then
it is the quantity $v_{\max}$ that determines the decoherence. In
the last two cases $\cK(t)$ is proportional to $t^2$, and one can apply the results of~\cite{Paladino2001}, substituting for $v$ either $v_{\min}$ or $v_{\max}$.

Since $r_{\scriptsize \textrm{opt}}$ depends on the temperature one can expect crossovers between different regimes as a function of temperature. A similar mesoscopic behavior of the decoherence rate was discussed for a microwave-irradiated Andreev interferometer~\cite{Lundin}.

Note that the result~(\ref{eq:0018b}) for the long-range interaction
cannot be reproduced by the Gaussian
approximation since in the latter case the decoherence would be determined by the nearest neighbors of the qubit. That can be seen from explicit expression for the phase accumulation following from the Gaussian model, see Eq.~(\ref{Gauss}) below. Being substituted in Eq.~(\ref{eq:Hotlsmark1}) this expression leads to divergence of the integral over $u$ at its upper limit that physically means dominating role of nearest neighbors of the qubit. At the same time, the spin-fluctuator model implies that the most important contributions are given by the fluctuators with $v(r) \sim \gamma_0$.

The above procedure still leaves unanswered a delicate question: Can
contributions of several \fs be described by averages over the
fluctuators' parameters?  The situation with a qubit interacting with environment in fact differs from that of a resonant two-level system in spin or phonon echo experiments. In the first case the experiment is conducted using a single qubit surrounded by a set of fluctuators
with fixed locations, while in the second case \emph{many}
resonant TLSs participate the absorption. Consequently, one can
assume that each TLS has its own environment and calculate the
properties averaged over positions and transition rates of the
surrounding fluctuators. How many surrounding fluctuators one
needs to replace the set of fluctuators with fixed locations (and
transition rates) by an averaged fluctuating medium? This issue was
studied numerically in~\cite{Galperin2004}. The analysis showed that
one needs really many ($\gtrsim 100$) fluctuators to avoid strong
mesoscopic fluctuations.

\section{Relevance to experiments}\label{experiments}

In this section we will briefly review some experiments were we believe that the theory of non-Gaussian noise is relevant. While there are some indications of non-Gaussian behavior,  to our knowledge, the existing experiments are not conclusive enough.
However, we should emphasize that observation of non-Gaussian effects was not among the goals of thee experiments. The main goal was to achieve the longest possible decoherence times. Probably as progress is made the situations where only one or a few independent
noise sources are important would become rather usual than exceptional making the non-Gaussian effects more pronounced. We also believe that experimental studies of the  non-Gaussian effects would provide useful information on the environmental degrees of freedom.
Therefore devices showing pronounced non-Gaussian behavior, while not necessarily being the optimal qubits, may serve as useful research tools.

\subsection{Plateaus in the echo signal}

\begin{figure}[h]
\begin{center}
\includegraphics[width=8.0cm]{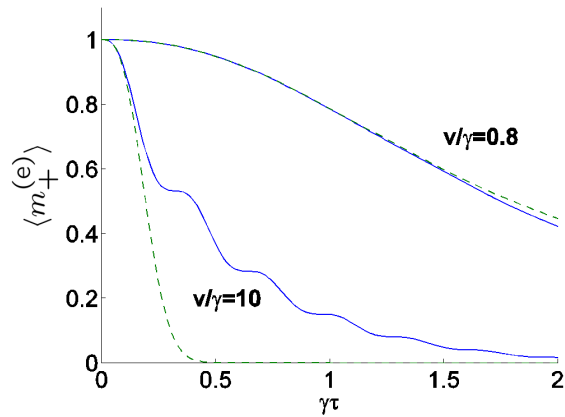}
\end{center}
\caption{Echo
signal for different values of the ratio $v/\gamma$ (shown near the
curves), Eq. (\ref{eq:04}). Dashed lines - calculations along the Gaussian
approximation, Eq. (\ref{Gauss}).
\label{fig:plateau}}
\end{figure}

Using Eqs.~(\ref{eq:gaussdec}) and (\ref{eq:t2g}) we have calculated the decay of the free induction signal due to a single fluctuator in the Gaussian approximation at large times. Let us return to
Eq.~(\ref{eq:theta2echo}) and evaluate the same for the echo signal,
but at arbitrary times. The result is
\begin{equation}
  \langle \psi^2\rangle
= \frac{v^2}{2\gamma^2}\left[4\gamma\tau_{12} -3 +4e^{-2\gamma \tau_{12}}
- e^{-4\gamma \tau_{12}} \right] \label{Gauss}
\end{equation}
with $\langle \me \rangle=e^{-\langle \psi^2\rangle/2}$ as before.
This should be compared with the exact result, Eq.~(\ref{eq:04}). In Fig.~\ref{fig:plateau} both results are shown for both a weakly ($v/\gamma=0.8$)
and a strongly ($v/\gamma=10$) coupled fluctuator.
As has been discussed before,  the Gaussian approximation is accurate for $v\lesssim\gamma$, while at $v>\gamma$ the two results are qualitatively different. In particular, the  plateaus in the time dependence of the echo signal shown in Fig.~\ref{fig:plateau} are beyond the Gaussian approximation. We believe that such a plateau
 was experimentally observed in Ref.~\cite{Nakamura02} (see
 Fig. 3 there). In the limit $v \gg \gamma, \sqrt{\gamma/\tau}$,
 equation \eref{eq:04} acquires a simple form:
 \begin{equation}
 \langle \me \rangle=e^{-2\gamma\tau}\left[1+\frac{\gamma}{v}\, \sin 2v \tau \right]\, .
\label{eq:plat}
 \end{equation}
According to Eq.~(\ref{eq:plat}), the plateau-like features
($d\langle \me \rangle/d\tau \approx 0$) occur at $v \tau \approx  k
\pi$ and their heights
$\langle \me \rangle\approx e^{-2\pi k\gamma/v}$ exponentially
decay  with the number $k$.
Measuring experimentally  the position and the height of the first
plateau, one  can determine both the fluctuator coupling strength $v$ and its switching rate $\gamma$.  For example, the echo signal
measured in Ref.~\cite{Nakamura02} shows
a plateau-like   feature at $\tau_{12}=3.5$~ns at the height
$\langle \me \rangle=0.3$, which yields $v\approx 143$~MHz,  and $\gamma \approx 27$~MHz.  If the fluctuator is  a charge trap near
the gates  producing a dipole electric field its coupling strength is $v=e^2({\bf a}\cdot {\bf r})/r^3$.  Using the gate-qubit distance $r\approx 0.5\ \mu$m, we obtain a reasonable estimate for the tunneling distance between the charge trap and the gate, $a\sim 20$ \AA. A more extensive discussion of this is found in Ref. \cite{Galperin2006}.

\subsection{Flux qubit}

In Sec. \ref{ensemble-average} we discussed a model with a broad distribution of coupling strengths, $v$. This is appropriate in a situation were the noise sources are distributed uniformly in space and act on the qubit via a long range (power law) force. In this section we want to apply the spin-fluctuator model to experiments on flux qubits~\cite{Yoshihara06,Galperin07}. Since the microscopic source of flux noise is not clarified (see, e.\,g.,~\cite{Harris08} and references therein), it is not clear what would be the most
reasonable distribution of $v$. We have adopted  the simplest model,
where the coupling parameters $v$ are  narrowly distributed around some characteristic value $\bar v$. In other words,  $v$ and $\gamma$ are supposed to be uncorrelated, $\cP(v,\gamma)=\cP_v(v)\cP_\gamma(\gamma)$, and
\begin{equation}
  \label{eq:pvpg}
  \cP_\gamma(\gamma)
   =\frac{P_0T}{\gamma}\varphi(\gamma_0-\gamma)\varphi(\gamma-\gamma_{\min}),
  \qquad
  \cP_v(v) = \delta(v-\bar{v}).
\end{equation}
Using Eq.~(\ref{eq:pvpg}) and the expressions for $\langle \me \rangle$ for either the Gaussian [Eq.~(\ref{Gauss})] or the spin-fluctuator [Eq.~(\ref{eq:04})] models we obtain the quantity $\cK$ defined in Eq.~(\ref{eq:avlog}).
In the Gaussian model:
\begin{equation}
   \label{eq:014}
  \cK_G(2\tau_{12}) =2 P_0T\bar{v}^2\tau_{12}^2 \times \left\{ \begin{array}{ccl}
2\gamma_0 \tau_{12}/3 &,& \gamma_0 \tau_{12} \ll 1 \\
\ln 2 &,& \gamma_0 \tau_{12} \gg 1 \, . \end{array} \right.
\end{equation}
We see that at long times, $\gamma_0 \tau_{12} \gg 1$, we have a quadratic dependence on time, which manifests a Gaussian decay of the echo signal. At short times, $\gamma_0 \tau_{12} \ll 1$, the result is multiplied by an additional factor $\gamma_0 \tau$, which is the probability for a single flip of the fastest fluctuators.

In the spin-fluctuator model there are two limiting cases:
\begin{itemize}
\item[(i)]
When $\bar{v} \ll \gamma_0$
 \begin{equation}
   \label{eq:018}
   \cK_{\mathrm{sf}}(2\tau_{12}) \approx
  \left\{\begin{array}{ccll}
     4P_0T\gamma_0\bar{v}^2  \tau_{12}^3/3 &,& &\tau_{12} \ll \gamma_0^{-1},  \\
     P_0T(2\ln 2)\, \bar{v}^2  \tau_{12}^2 &,&   \gamma_0^{-1}\ll &\tau_{12}\ll\bar{v}^{-1},\\
     P_0T\alpha\bar{v} \tau_{12} &,&   \bar{v}^{-1}\ll &\tau_{12}.
  \end{array} \right.
\end{equation}
where $\alpha\approx 3$.  At small times $\tau_{12}\ll\bar{v}$ we arrive at the same result as in the Gaussian approach, Eq.~(\ref{eq:014}). However, at large times, $\tau_{12}\gg\bar{v}^{-1}$, the exact result dramatically differs from the prediction of the Gaussian approximation. To understand the origin of the non-Gaussian behavior notice that for $\cP_\gamma (\gamma) \propto 1/\gamma$, the decoherence is dominated by the fluctuators with $\gamma \approx v$.
Indeed, very ``slow'' fluctuators produce slow varying fields, which are effectively refocused in course of the echo experiment. As to the ``too fast'' fluctuators, their influence is reduced due to the effect of motional narrowing. We have already learned that only the fluctuators with $v \ll \gamma$ produce Gaussian noise. Consequently, the noise in this case is essentially non-Gaussian. Only at times $\tau_{12}\ll\bar{v}^{-1}$, which are too short for these most important fluctuators to switch, the decoherence is dominated by the faster fluctuators contribute, and the Gaussian approximation turns to be valid.

\item[(ii)]
When $\bar{v} \gg \gamma_0$ we find
\begin{equation}
  \label{eq:018b}
    \cK_{\mathrm{sf}}(2\tau_{12}) \approx
  \left\{\begin{array}{ccl}
     4P_0T\gamma_0 \bar{v}^2 \tau_{12}^3/3 &,& \tau_{12} \ll  \bar{v}^{-1},  \\
    2 P_0T\gamma_0  \tau_{12} &,&    \tau_{12}\gg \bar{v}^{-1}.
  \end{array} \right.
\end{equation}
In this case \textit{all} fluctuators are strongly coupled to the qubit. Therefore, the long-time decoherence is essentially non-Gaussian.
\end{itemize}

\begin{figure}[h]
\begin{center}
\includegraphics[width=9.0cm]{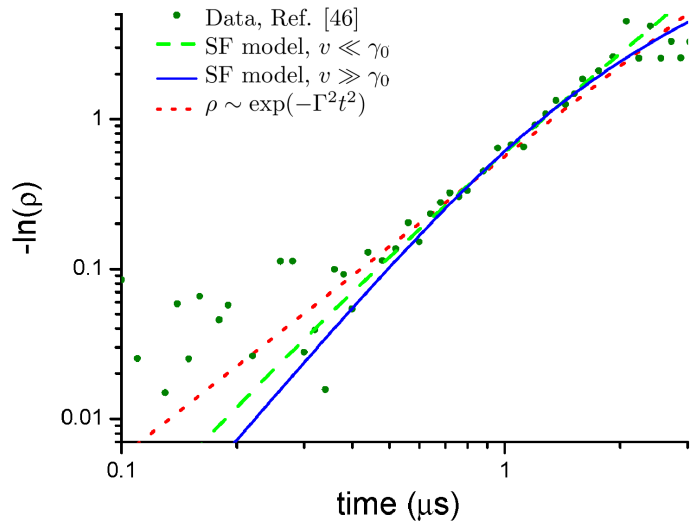}
\end{center}
\caption{ Fit of the experimental data from \cite{Yoshihara06} to the
Gaussian model, and to the spin-fluctuator model with $\bar{v}\ll\gamma_0$
[Eq. (\ref{eq:018})] or $\bar{v}\gg\gamma_0$ [Eq. (\ref{eq:018b})].
\label{fig:yoshi}}
\end{figure}

We thus conclude that it is the long-time decoherence that is most sensitive to the particular model of the noise. Unfortunately, at long times the signal usually is weak and obscured by noise. One of the possible ways to experimentally identify the non-Gaussian behavior is based on the fact that the typical fluctuator strength, $\bar{v}$, enters the the expressions for the Gaussian and non-Gaussian decay in different ways. In particular, according to the Gaussian approximation, $\cK_m(t) \propto \bar{v}^2$ at all times. Contrarily, according to the SF model [Eqs.~(\ref{eq:018}), (\ref{eq:018b})] the powers of $\bar{v}$ are different for different times. By fitting to the spin-fluctuator model this parameter can be extracted, and from this we can infer the flux change corresponding to the flip of a single fluctuator. We have performed such fits to the experimental data of \cite{Yoshihara06}, the details are given in \cite{Galperin07}, and one example is reproduced in Fig.~\ref{fig:yoshi}.
As can be seen, it is difficult to determine on the basis of these experiments whether non-Gaussian effects are important, or the Gaussian approximation is sufficient, both models fit the data equally well. However, {\it if} we choose to fit the spin-fluctuator model and fit not only to one experimental curve, but to the whole set of curves for different working points of the qubit, we find that Eq.~(\ref{eq:018}) for the case $\bar{v} \ll \gamma_0$
provides the better overall fit. The change of flux due to a single
fluctuator flip is $\lesssim 10^{-5}\Phi_0$, where $\Phi_0$ is the
flux quantum.

\section{Microscopic sources of telegraph noise}\label{microscopic}
In this section we briefly discuss of microscopic noise
sources that can produce classical telegraph noise.

\paragraph{Charge noise --}
The obvious source of RTP charge noise is a charge, which jumps
between two different locations in space. Less clear is where these charges are actually located and what are the two states.
The first attempt of constructing such a model in relation to qubit decoherence appeared in \cite{Paladino2001}, where electrons tunneling between a localized state in the insulator and a metallic gate was studied. This model has been further studied in \cite{Lerner,Abel08}. Later, experimental results \cite{Astafiev04} indicated a linear dependence of the relaxation rate on the energy splitting of the two qubit states. One also has to take into account that in the experimental setup there is no normal metal in the vicinity of the qubit: all gates and leads should be in the superconducting state at the temperatures of experiment. These two facts suggest that the model~\cite{Paladino2001} is irrelevant for the decoherence in charge qubits~\cite{Nakamura02} and favored a model with superconducting electrodes \cite{Faoro05}.
\begin{figure}[h]
\begin{center}
\includegraphics[width=8.0cm]{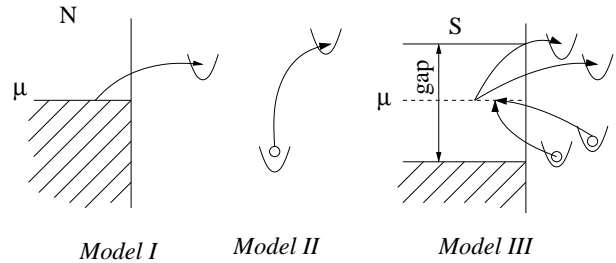}
\end{center}
\caption{ Three possible models for the fluctuating charges: Model I, electrons jumping between a localized state and a normal metal, as
discussed in \cite{Paladino2001,Lerner,Abel08}. Model II, electrons
jumping between localized states. Model III, electrons jumping between localized states and a superconductor, as discussed in \cite{Faoro05}.
\label{fig:models}}
\end{figure}
In this model, the two electrons of a Cooper pair are split and tunnel separately to  some localized states in the insulator (see
Fig.~\ref{fig:models} for an illustration of this (Model III) and
other models). A constant density of these localized states
gives a linearly increasing density of occupied pairs, in
agreement with experiments \cite{Astafiev04}.  This model was
criticized \cite{Faoro06} because it required an unreasonably high
concentration of localized states, and a more elaborate model was
proposed. However, it was shown~\cite{Lerner,Abel08} that allowance
for quantum effects of hybridization between the electronic states
localized at the traps and extended states in the electrodes relaxes
the above requirement. At present it seems that no solid conclusions
can be drawn based on the available experiments.

\paragraph{Noise of critical current --}
The microscopic mechanism and the source of the fluctuations on the critical current in a Josephson junction is a long-standing open
problem. These fluctuations were initially attributed to the charges
tunneling or hopping between different localized states in the
barrier forming glass-like TLSs. However, a more detailed comparison with experiment revealed an important problem -- the experimentally observed noise spectrum~\cite{Wellstood04}  was proportional to $T^2$ that is incompatible with the assumption of constant TLS density of
states. The interest in the critical current fluctuations was recently renewed because of their importance for superconducting qubits. The new experiments~\cite{Eroms06} on fluctuations in small Al junctions -- similar to those used in several types of qubits~\cite{Chiorescu03} -- in normal state brought a new puzzle. It turned out that the temperature dependence of the noise power spectrum in the normal state is \textit{linear}, and the noise power is much less than that reported for large superconducting contacts. A plausible explanation of such behaviors is given in~\cite{Faoro07} where it is suggested that the critical current noise is due to electron trapping in shallow subgap states that might be formed at the superconductor-insulator boundary. This mechanism is similar to that suggested earlier for the charge noise~\cite{Faoro06}.

\paragraph{Flux noise  --}
Studies of the flux noise in superconducting structures
have a long history. As early as in the 1980s it was demonstrated
it is the flux and not the critical current noise that limits the
sensitivity of dc SQUIDs~\cite{Koch83}. The interest in this problem
was recently renewed when it was realized that flux noise can limit
the coherence in flux and phase superconducting
qubits~\cite{Yoshihara06,Martinis03}. Two recent models for
fluctuators producing low-frequency noise were suggested. The first
one~\cite{Koch07} attributes the flux noise to the electron hopping
between traps in which their spins have fixed, random orientations. The second one~\cite{Sousa07} proposes that electrons flip their spins due to interaction with tunneling TLSs and phonons.
These models were recently criticized in Ref.~\cite{Faoro08} where it was stated that it is hard to justify the assumptions behind both
models. In that paper, a novel mechanism - based on spin diffusion
along the surface of a superconductor - was suggested. This model
seems to agree with recent experiments~\cite{Bialczak07} on
measurements of the $1/f$ noise. It remains to be understood whether the surface spin fluctuations lead to pronounced non-Gaussian behavior in the qubit decoherence.

\section{Conclusions}

In this review we have discussed the spin-fluctuator model for qubit
dephasing. This model provides a simple, solvable, yet in many situations realistic model of the qubit environment. In particular, we show how to apply the model in situations where the noise shows $1/f$ behavior. The model shows pronounced non-Gaussian behavior, and thus serves as an example of possible deviations from the Gaussian approximation, as well as shedding light on the limitations of the Gaussian approximation.

The main results obtained in studying this model can be summarized
as:
\begin{itemize}
\item{A single fluctuator, characterized by the coupling strength $v$ and the switching rate $\gamma$ can be classified as weak ($v<\gamma$) or strong ($v>\gamma$). For weak fluctuators the spin-fluctuator model reproduces the Gaussian result, whereas for strong fluctuators it shows non-Gaussian behavior}

\item{This non-Gaussian behavior persists even in the limit of long time and for an arbitrary number of independent fluctuators, as long as each fluctuator is strong. This can be understood as a consequence of the $\delta$-functions at the extreme of the phase distribution function.}

\item{In the non-Gaussian case, the time correlation function of the noise is not sufficient to determine the qubit dephasing and a more detailed model must be specified.}

\item{In the non-Gaussian case, individual fluctuators leave signatures in the measured signal, e.\,g., plateaus in the echo signal, that can be used to identify the fluctuator parameters.}
\end{itemize}

We should emphasize that a strong fluctuator, giving non-Gaussian effects, does not imply strong decoherence and therefore a bad qubit performance. It indicates only that the coupling is strong relative to the switching rate. However, independently of the performance of the device as  a working qubit, i. e., on whether it achieves a long dephasing time, we believe that searching for signatures of non-Gaussian behavior can provide valuable information on the nature of the noise source. Given the present uncertain state of understanding of the microscopic sources of noise in most solid state qubit devices, this seems an important undertaking, and analyzing experiments according to the formulas we have presented can be a useful tool in this process.

\section*{Acknowledgments}

We are thankful to Norwegian Research Council for financial support
along the STORFORSK and BILAT programs, as well as to NEC-Laboratories America, Argonne National
Laboratory, USA,  and National Center for Theoretical Sciences of
R.O.C, Taiwan for partial financial support and hospitality.


\section*{References}
\begin{thebibliography}{99}
\bibitem{Dutta81} P. Dutta and P.M. Horn,
  Rev. Mod. Phys. \textbf{53}, 497 (1981).

\bibitem{Weissman88} M.B. Weissman, Rev. Mod. Phys. \textbf{60}, 537
  (1988).

\bibitem{Zorin96} A.B. Zorin, F.J. Ahlers, J. Niemeyer, T. Weimann,
  H. Wolf, V.A. Krupenin, and S.V. Lotkhov, Phys. Rev. B
  \textbf{53}, 13682 (1996).

\bibitem{Astafiev04} O. Astafiev, Y.A. Pashkin, Y. Nakamura,
  T. Yamamoto, and J.S. Tsai, Phys. Rev. Lett. \textbf{93}, 267007
  (2004).

\bibitem{Astafiev06} O. Astafiev, Y.A. Pashkin, Y. Nakamura,
  T. Yamamoto, and J.S. Tsai, Phys. Rev. Lett. \textbf{96}, 137001
  (2006).

\bibitem{Kogan1996} Sh.M. Kogan, \textit{Electronic Noise and
    Fluctuations in Solids}, Cambridge University Press, Cambridge,
  UK, (1996).

\bibitem{RTP} M.J. Buckingham, \textit{Noise in Electronic Devices and
Systems} (Ellis Horwood Ltd., New York, 1983); M.J. Kirton and
M.J. Uren, Adv. Phys. \textbf{38}, 367 (1989).

\bibitem{RTN} C.E. Parman, N.E. Israeloff, and J. Kakalios, \prb
  \textbf{44} 8391 (1991); K.S. Ralls, W.J.~Skocpol, L.D. Jackel,
  R.E. Howard, L.A.  Fetter, R.W. Epworth, and D.M. Tennent, Phys. Rev. Lett.
\textbf{52}, 228 (1984); C.T. Rogers and R.A. Buhrman,
Phys. Rev. Lett. \textbf{53}, 1272 (1984) and \prl \textbf{55}, 859
  (1985);
T.~Duty, D. Gunnarsson, K. Bladh, and P. Delsing, \prb
  \textbf{69} 140504(R) (2004); M. Peters, J. Dijkhuis, and
  L. Molenkamp, J. Appl. Phys. \textbf{86}, 1523 (1999); J. Eroms, L. van
Schaarenburg, E. Driessen, J. Plantenberg, K. Huizinga, R.
Schouten, A. Verbruggen, C.~Harmans, and J.~Mooij, Appl.
Phys. Lett. \textbf{89}, 122516 (2006).

\bibitem{Simmonds04}
R. W. Simmonds, K.M. Lang, D.A.~Hite, S. Nam, D. P. Pappas,
and J.M. Martinis, \prl \textbf{93}, 077003 (2004); K. B.
Cooper, M. Steffen, R. McDermott, R W. Simmonds, S. Oh, D.A.~Hite, D.P. Pappas, and J.M. Martinis, ibid. \textbf{93}, 180401
(2004).

\bibitem{Ithier05} G. Ithier, E. Collin, P. Joyez, P.J. Meeson,
D. Vion, D. Esteve, F. Chiarello, A. Shnirman, Y.~Makhlin,
J. Schriefl, and G. Sch\"on, Phys. Rev. B \textbf{72}, 134519 (2005)

\bibitem{Shnirman05}
A. Shnirman, G. Sch\"on, I. Martin, and Y. Makhlin, Phys. Rev.
Lett. \textbf{94}, 127002 (2005).

\bibitem{KlauderAnderson} R. Klauder and
P.W. Anderson, Phys. \, Rev.  {\bf   125}, 912 (1962).

\bibitem{HuWalker} P. Hu and L. Walker, Solid State Commun.
{\bf   24}, 813 (1997);
R. Maynard, R. Rammal, and R.~Suchail, J. \,
  Phys. (Paris) \ Lett. {\bf 41}, L291 (1980).

\bibitem{BlackHalperin} J.L. Black and B.I. Halperin, \prb {\bf
16}, 2879 (1977).

\bibitem{SMS} W.E. Moerner, Science {\bf 265}, 46 (1994); W.E.
Moerner, M. Orrit, Science {\bf 283}, 1670 (1999); E.~Geva, P.D.
Reily, J.L. Skinner, Acc. \ Chem. \ Res. {\bf 29}, 579 (1996); E.
Barkai, Y. Jung, R. Silbey, \prl {\bf 87}, 207403 (2001).


\bibitem{Ludviksson1984} A.~Ludviksson, R.~Kree, and A.~Schmid,
  Phys. Rev. Lett. \textbf{52}, 950 (1984).

\bibitem{Kogan1984} S.M. Kogan and K.E. Nagaev,
Solid State Commun. \textbf{49}, 87 (1984).

\bibitem{Kozub1984} V.I.~Kozub, Sov. Phys. JETP \textbf{59},
  1303 (1984).

\bibitem{Galperin1991} Y.M. Galperin and
  V.L. Gurevich, Phys. Rev. B \textbf{43}, 1290 (1991).

\bibitem{Galperin1994}
Y.M. Galperin, N.~Zou, and K.A.~Chao, Phys. Rev B \textbf{49}, 13728 (1994).

  \bibitem{Hessling1995} J.P.~Hessling and Y.M.~Galperin,  Phys. Rev B
  \textbf{52}, 5082 (1995).

\bibitem{Paladino2001} E.~Paladino, L.~Faoro, G.~Falci. and R.~Fazio,
\prl \textbf{88}, 228304 (2002).

\bibitem{Paladino2003}  E.~Paladino, L.~Faoro, A.~D`Arriogo,
  A.~Mastelone, and G.~Falci,
Physica E (Amsterdam) \textbf{18}, 29 (2003).

\bibitem{Falci2003} G.~Falci, E.~Paladino, and R.~Fazio,
in \textit{ Quantum Phenomena in Mesoscopic Systems}
  edited by B.L. Altshuler and V. Tognetti  (IOS Press,
  Amsterdam, 2003.

\bibitem{Galperin2004} Y.M. Galperin, B.L. Altshuler, and D.V. Shantsev,
in \textit{Fundamental Problems
  of Mesoscopic Physics} edited by I.V.~Lerner et al.
  (Kluwer Academic Publishers, The Netherlands, 2004), pp.~141--165.

\bibitem{Falci2004} G.~Falci, A. D`Arriogo, A. Masteloni, E.~Paladino,
  and R.~Fazio,  \pra \textbf{70}, 040101 (2004).

\bibitem{Falci2005}  G.~Falci, A. D`Arriogo, A. Masteloni,
  E.~Paladino, \prl \textbf{94}, 167002 (2005).

\bibitem{Galperin2006} Y.M. Galperin, B.L. Altshuler, J. Bergli, and
  D.V.~Shantsev, \prl \textbf{96}, 097009 (2006).

\bibitem{Martin2006} I. Martin, Y.M.~Galperin, \prb \textbf{73},
  18020 (2006).

\bibitem{Bergli2006} J. Bergli, Y.M. Galperin, B.L. Altshuler, \prb
  \textbf{74}, 024509 (2006).

\bibitem{DiVincenzo2005} D.P.~DiVincenzo, D.~Loss, \prb \textbf{71},
  035318 (2005); R.M. Lutchyn, \L. Cywi{\'nski}, C.P.~Nave, and S. Das
  Sarma, \prb \textbf{78}, 0345 (2008); W.A. Coish, J. Fischer, and
  D. Loss, \prb \textbf{77}, 125329 (2008).

\bibitem{Galperin05a} Y.M. Galperin, D.V. Shantsev, J. Bergli and
  B.L. Altshuler, Europhys. Lett. \textbf{71}, 21 (2005).

\bibitem{Paladino08} E. Paladino, M. Sassetti, G. Falci, and U. Weiss,
  \prb \textbf{77}, 041303(R) (2008).

\bibitem{free-ind} Charles P. Slichter, ``Principles of Magnetic
  Resonance'', Springer series in Solid-State Sciences, Vol. 1, ed. by
  M. Cardona and H.J. Queisser, Springer-Verlag, Berlin (1978).


\bibitem{AstafievSS} O. Astafiev, Yu. A. Pashkin, T. Yamamoto,
  Y. Nakamura, and J.S. Tsai, Phys. Rev. B {\bf 69}, 180507(R) (2004).

\bibitem{Mims} W.B. Mims, in \emph{Electron Paramagnetic Resonance},
  edited by S. Geschwind (Plenum, New York, 1972).

\bibitem{Itakura03} T.~Itakura and Y.~Tokura, Phys. Rev. B \textbf{67},
  195320 (2003).

\bibitem{TLS} P.W. Anderson, B.I. Halperin, and C.M. Varma, Philos.
Mag. \textbf{25}, 1 (1972); W.A. Phillips, J. Low
Temp. Phys. \textbf{7}, 351 (1972).

\bibitem{gamma} J. J\"ackle, Z. Phys. \textbf{257}, 212 (1972).

\bibitem{Black} J.L. Black and B.L. Gyorffy, \prl {\bf 41}, 1595
    (1978); J.L. Black, in \textit{Glassy Metals, Ionic Structure,
    Electronic Transport and Crystallization} (Springer, New York,
1981).



\bibitem{dist2} B.I. Halperin, Ann. N. Y. Acad. Sci. \textbf{279}, 173 (1976);
J.L. Black, Phys. Rev. B \textbf{17}, 2740 (1978).

\bibitem{Laikhtman} B.D. Laikhtman, Phys. Rev. B {\bf
    31}, 490 (1985).

\bibitem{Chandrasekhar} S. Chandrasekhar, Rev. Mod. Phys. {\bf 15},1
  (1943).

\bibitem{Lundin} N. I. Lundin and Y. M. Galperin, \prb {\bf 63},
  094505 (2001).

\bibitem{Nakamura02} Y. Nakamura, Yu.A. Pashkin, T. Yamamoto, and J.S. Tsai,
Phys. Rev. Lett.{\bf  88}, 047901 (2002).

\bibitem{Yoshihara06} F. Yoshihara, K. Harrabi, A.O. Niskainen,
  Y. Nakamura, and J.S. Tsai, Phys. Rev. Lett. {\bf 97}, 167001
  (2006).

\bibitem{Galperin07} Y.M. Galperin, B.L. Altshuler, J. Bergli,
D. Shantsev, and V. Vinokur,  Phys. Rev. B {\bf 76}, 064531 (2007).

\bibitem{Harris08} R. Harris, M.W.~Johnson, S.~Han, A.J.~Berkley,
J.~Johansson, P.~Bunyk, E.~Ladizinsky, S.~Govorkov, M.C.~Thom,
S.~Uchaikin, B.~Bumble, A.~Fung, A.~Kaul, A. Kleinsasser,
M.H.S.~Amin, and D.V.~Averin, \prl \textbf{101}, 117003 (2008).

\bibitem{Lerner} A. Grishin, I. V. Yurkevich, and I. V. Lerner,
Phys. Rev. B {\bf 72}, 060509 (2005).

\bibitem{Abel08} B. Abel and F. Marquardt, arXiv:0805.0962 (2008).

\bibitem{Faoro05} L. Faoro, J. Bergli, B. L. Altshuler, and Yu. M. Galperin
Phys. Rev. Lett. {\bf 95}, 046805 (2005).

\bibitem{Faoro06} L. Faoro and L.B. Ioffe, \prl \textbf{96}, 047001 (2006).

\bibitem{Wellstood04} F.C. Wellstood, C. Urbina, and J. Clarke,
  Appl. Phys. Lett. \textbf{85}, 5296 (2004); D.J. van Harlingen,
  T.L. Robertson, B.L.T. Plourde, P.A. Reichardt, T.A. Crane, and
  J. Clarke, \prb \textbf{70}, 064517 (2004).

\bibitem{Eroms06} J. Eroms, L.C. van Schaarenburg, E.F.C. Driessen,
  J.H. Plantenberg, C.M. Huizinga, R.N.~Schouten, A.H. Verbruggen,
  C.J.P.M. Harmans, and J.E. Mooij, Appl. Phys. Lett. \textbf{89},
  122516 (2006).

\bibitem{Chiorescu03} I. Chiorescu, Y. Nakamura, C.J.P.M. Harmans,
  and J.E. Mooij, Science \textit{299}, 1869 (2003); J.M.~Martinis,
  S. Nam, J. Aumentado, and C. Urbina, \prl \textbf{89}, 117901
  (2002); D. Vion, A. Aassime, A. Cottet, P. Joyez, H. Pothier,
  C. Urbina, D. Esteve, and M. H. Devoret, Science \textbf{96}, 886
  (2002).

\bibitem{Faoro07} L. Faoro and L.B. Ioffe, \prb \textbf{75}, 132505 (2007).

\bibitem{Koch83} R.H.~Koch, J.~ Clarke, W.M.~Goubau, J.M.~Martinis,
  C.M.~Pegrum and D.J.~van~Harlingen, J.~Low. Temp. Phys. \textbf{51},
  207 (1983); F.C.~Wellstood, C.~Urbina, and J.~Clarke,
  Appl. Phys. Lett. \textbf{50}, 772 (1987).

\bibitem{Martinis03} J.M.~Martinis,  S.~Nam, J.~Aumentado, and
  K.~M.~Lang, \prb \textbf{67}, 094510 (2003).

\bibitem{Koch07} R.H.~Koch, D.P.~DiVincenzo, and J.~Clarke, \prl
  \textbf{98}, 267003 (2007).

\bibitem{Sousa07} R.~de Sousa, \prb \textbf{76}, 245306 (2007).

\bibitem{Faoro08} L. Faoro and L.B.~Ioffe, \prl \textbf{100}, 227005
  (2008).

\bibitem{Bialczak07} R.C. Bialczak, R. McDermot, M. Ansmann, H. Hofheinz,
  N. Katz, E. Lucero, M. Neeley,  A.D.~O'Connel, H. Wang,
  A.N. Cleland, and J.M.~Martinis, \prl \textbf{99}, 187006 (2007);
  S. Sendelbach, D. Hover, A. Kittel,  M. M\"uck,  J. M. Martinis,
  and R. McDermott, \prl \textbf{100}, 227006 (2008).


\end{thebibliography}
\end{document}